\documentclass[sigconf]{acmart}

\usepackage{bbm}
\usepackage{amsmath}
\usepackage{multirow}

\usepackage{algorithm}

\usepackage{newfloat}
\usepackage{listings}
\usepackage{xcolor}
\usepackage{colortbl}
\usepackage{booktabs}
\usepackage{CJKutf8}
\DeclareMathOperator*{\argmax}{argmax}
\DeclareMathOperator*{\argmin}{argmin}

\usepackage[noend]{algpseudocode}
\usepackage{algorithmicx,algorithm}

\newcommand{\eqdef}{\overset{\mathrm{def}}{=\joinrel=}}

\AtBeginDocument{%
  \providecommand\BibTeX{{%
    \normalfont B\kern-0.5em{\scshape i\kern-0.25em b}\kern-0.8em\TeX}}}

\setcopyright{acmcopyright}
\copyrightyear{2022}
\acmYear{2022}
\acmDOI{10.1145/123456.123456}

\acmConference[KDD '22]{KDD '22: SIGKDD Conference on Knowledge Discovery and Data Mining}{August 14-18, 2022}{Washington D.C.}
\acmBooktitle{KDD '12: ACM SIGKDD Conference on Knowledge Discovery and Data Mining, August 14-18, 2022, Washington D.C.}
\acmPrice{15.00}
\acmISBN{978-1-4503-XXXX-X/18/06}

\begin{document}

\title{Adversarial Gradient Driven Exploration for Deep Click-Through Rate Prediction}

\author{Kailun Wu, Zhangming Chan, Weijie Bian, Lejian Ren, \\ Shiming Xiang$^{*}$, Shuguang Han, Hongbo Deng, Bo Zheng}
\affiliation{
\institution{Alibaba Group \quad  * Institute of Automation, Chinese Academy of Sciences}
\city{Beijing}
\country{People's Republic of China}
}

\email{{kailun.wukailun, zhangming.czm, weijie.bwj, lejian.rlj}@alibaba-inc.com}
\email{{shuguang.sh, dhb167148, bozheng}@alibaba-inc.com, smxiang@nlpr.ia.ac.cn} 

\renewcommand{\shortauthors}{Wu and Chan et al.}
\renewcommand{\authors}{Kailun Wu, Zhangming Chan, Weijie Bian, Lejian Ren, Shiming Xiang, Shuguang Han, Hongbo Deng, Bo Zheng}

\begin{abstract}
Exploration-Exploitation (E{\&}E) algorithms are commonly adopted to deal with the feedback-loop issue in large-scale online recommender systems. Most of existing studies believe that high uncertainty can be a good indicator of potential reward, and thus primarily focus on the estimation of model uncertainty. We argue that such an approach overlooks the subsequent effect of exploration on model training. From the perspective of online learning, the adoption of an exploration strategy would also affect the collecting of training data, which further influences model learning. To understand the interaction between exploration and training, we design a Pseudo-Exploration module that simulates the model updating process after a certain item is explored and the corresponding feedback is received. We further show that such a process is equivalent to adding an adversarial perturbation to the model input, and thereby name our proposed approach as an the Adversarial Gradient Driven Exploration (AGE). For production deployment, we propose a dynamic gating unit to pre-determine the utility of an exploration. This enables us to utilize the limited amount of resources for exploration, and avoid wasting pageview resources on ineffective exploration. The effectiveness of AGE was firstly examined through an extensive number of ablation studies on an academic dataset. Meanwhile, AGE has also been deployed to one of the world-leading display advertising platforms, and we observe significant improvements on various top-line evaluation metrics.
\end{abstract}

\ccsdesc[500]{Information systems~Recommender systems}
\ccsdesc[500]{Information systems~Display advertising}

\keywords{Exploration and Exploitation, Recommender Systems, Click-Through Rate Prediction, Online Advertising}

\maketitle


\section{Introduction}
 
Click-through Rate (CTR) prediction is the core module for many online recommendation systems. 
While receiving a user request, a recommender system usually retrieves a set of candidate items, ranks them, often by the predicted likelihood of user click, and finally displays to end users. 
Recent progress on deep neural networks expedites the development of CTR prediction techniques. 
A variety of deep neural predictive models have been proposed and widely adopted in various large-scale industrial applications such as movie recommendation systems, e-commerce platforms, and online advertising platforms~\cite{cheng2016wide,guo2017deepfm,zhou2018deep,zhou2019deep,pi2019practice,chan2020selection,li2019multi,feng2019deep,bian2022can,pi2020search}. 

As the de facto standard, CTR models are commonly trained on top of the collected impression data. After being deployed online, such a model produces a new stream of impression data, which will then be used for model updating. This creates the so-called feedback-loop issue~\cite{sculley2015hidden,zhang2020retrain}, and the exposure bias will be gradually amplified, resulting in strong Matthew effects in recommender systems~\cite{chen2020bias}. The direct consequence is that new and long-tailed items can barely break the loop and grow successfully, as the model predicts them with less certainty~\cite{zeldes2017deep,xu2022ukd}. With subpar model performance for those items, a recommender system may redirect users to uninterested items, causing less user engagement.

\begin{figure}[!t]
	\centering
	\includegraphics[width=1\linewidth]{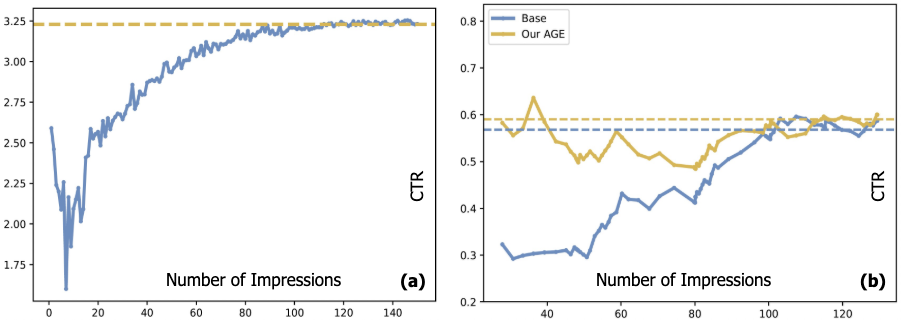}
	\caption{The change of click-through rate with the increase of impressions for each item. Figure (a) illustrates the CTR convergence curve for the current production model (average over items); Figure (b) compares the CTR convergence curves between the production model and our proposed method (average over a set of selected popular items).}
	\label{fig:ctr_cs}
	\vspace{-1em}
\end{figure}

To understand how the model predictions can be affected by the amount of impressions, we choose a list of items with more than 14,000 impressions in our production system, and monitor the change of click-through rates with the increase of impressions for those items. Our production system is one of the leading displaying advertisement platforms in the world. Specifically, as illustrated by Figure~\ref{fig:ctr_cs}, we plot the true click-through rate over the number of impressions received by each item. It appears that a new item in our system requires an average of 10,000 impressions in order to reach convergence. This introduces the common dilemma for many online systems -- how to redirect users to the most interesting items, often with an abundant number of impressions already (and better prediction accuracy), while reserving sufficient impressions for new and long-tailed items at the same time.

Algorithms fall under the exploration-exploitation (E{\&}E) framework are often adopted to resolve the above problem~\cite{auer2002using,chapelle2011empirical,liquin2017explain}. In recommendation systems, common approaches such as the contextual multi-armed bandit~\cite{li2010contextual,li2011unbiased} models this problem as follows. At each step, the system selects an action (recommends an item $i$ to a user) based on a policy $\text{P}$. With the goal of maximizing the cumulative reward (often measured by the total number of clicks) over time, the policy leverages the exploitation of items with high estimated reward $\mu_i$ (based on current knowledge) with the exploration of items with high uncertainty of the reward $\delta_i$. After recommendation, the system will receive the true reward (e.g. click) for policy updating. The overall process can be briefly summarized as Formula \ref{eq:ee-framework}. Here, $pctr'_i$ stands for the ranking score for item $i$, and the function $k(\cdot)$ indicates the trade-off strategy. UCB-like approaches~\cite{auer2002using,li2010contextual} usually adopt the upper bound of potential reward, whereas Thompson Sampling-like methods~\cite{chapelle2011empirical} choose an action through sampling from the estimated probability distribution.

\begin{equation}
\label{eq:ee-framework}
 \text{Policy P:}\quad {pctr}'_i \eqdef
 k\big(\mu_{i}, \delta_{i}\big)
\end{equation}

Previous studies often believe that high uncertainly is a good indicator of potential reward. Accordingly, uncertainty estimation has become the core module for many E{\&}E algorithms. Uncertainty may originate from data variability, measurement noise and model unstableness (e.g., parameter randomness)~\cite{zeldes2017deep}. Existing research primarily focused on estimating model uncertainty, and typical approaches include the Monte Carlo Dropout~\cite{gal2016dropout}, Bayesian Neural Networks for weight uncertainty~\cite{blundell2015weight}, Gaussian process for prediction uncertainty~\cite{du2021exploration,rasmussen2003gaussian}, and gradient norm (of model weights) based uncertainty modeling~\cite{song2021underestimation,zhang2020neural}.

We argue that the above assumption does not provide a holistic view for exploration. For data-driven online systems, the ultimate benefit of exploration comes from the feedback information acquired from the exploration process, and the further model update based on such data. Whereas the uncertainty itself cannot completely reflect such a whole process. To this end, we introduce a \textbf{Pseudo-Exploration} module to simulate model training after a certain item is explored and the corresponding feedback is received. Later on, we discover that an effective exploration action should be determined not only by the prediction uncertainty but also by the direction of exploration that leads to the maximal change of prediction output (i.e., the gradient). Further analysis also shows that this process is equivalent to adding an adversarial perturbation to the input feature; therefore, we name this approach as the \textbf{A}dversarial \textbf{G}radient based \textbf{E}xploration, \textbf{AGE} for short.

There are two important distinctions between the traditional E\&E algorithms and AGE. Firstly, AGE redefines the goal of an exploration as seeking for the exploration actions that can facilitate a faster model convergence. This differs from most of the previous studies in which the utility of an exploration is solely determined by the model uncertainty. Secondly, instead of a direct combination of the uncertainty score and the prediction score, as did in many previous studies~\cite{du2021exploration,song2021underestimation,zhang2020neural,auer2002using}, AGE transforms the exploration problem into the injection of adversarial perturbation to the input. This often results in an improved model robustness~\cite{rozsa2016accuracy}.

Furthermore, we discover that not all of the items are worth exploring in industrial systems. In the conventional top-K recommendation paradigm, only a small number of items can be finally displayed to end users. Items with extremely low click-through rates, despite having high model uncertainty, are still with less value for exploring. With an extensive amount of exploration, we may acquire more accurate predictions for those items; however, because of the noncompetitive prediction scores, they still cannot be displayed in the post-exploration stage. For this reason, we propose a dynamic gating unit to pre-determine the usefulness of an exploration action. In this paper, we experiment a simple heuristic -- we conduct an exploration if the prediction score is higher than the item-level average of click-through rate.

To summarize, our main contributions are listed as follows:

\begin{itemize}
\item Different from the majority of Exploration-Exploitation algorithms that concentrate on estimating model uncertainty, we propose to measure the utility of an exploration based on its influence on subsequent model training, and thus design a Pseudo-Exploration module to simulate model updating after the exploration. This provides a new perspective for defining the utility of an exploration.

\item We discover that the above pseudo-exploration process is essentially an injection of adversarial perturbation to model input. For this reason, we propose a novel, Adversarial Gradient based Exploration (AGE) algorithm for handling the E\&E problem for recommendation. In addition, AGE introduces a Dynamic Gating Unit to pre-filter the items with limited value for exploration.

\item We validate the effectiveness of AGE with an academic dataset and further examine its performance through online A/B testing on Alibaba display advertising system. AGE exhibits superior performances on several top-line metrics, and a significant acceleration of model convergence.
\end{itemize}

In the below sections, we first survey the related work in Section \ref{relate}, and then introduce the details of our proposed AGE algorithm in Section \ref{sec:model}. With a comprehensive description of the dataset and evaluation metrics in Section~ref{exp}, we further evaluate the effectiveness of AGE through an extensive number of experiments using both academic datasets and online A/B testing in Section \ref{sec:aca_exp}.

\section{Related work}
\label{relate}
The problem of Exploration-Exploitation (E\&E) trade-off is a long-standing research issue in the machine learning community, and a plentiful of related approaches have been proposed and examined in various settings~\cite{li2010contextual,li2011unbiased,balabanovic1998exploring,vanchinathan2014explore,mcinerney2018explore,nguyen2019recommendation,ban2021multi,zhang2020neural,song2021underestimation,du2021exploration,guo2020deep}.

\subsection{Exploration-Exploitation Trade-off} 
Multi-Armed Bandit (MAB) is a typical approach for dealing with the E\&E problem~\cite{auer2002nonstochastic,abe2003reinforcement,dani2008stochastic}. In MAB, we usually have a set of available arms, and in each round, we select one of them to play based on the trade-off between the current reward and the potential reward. The two types of rewards can be estimated in more accurately with an increasing number of selections of the corresponding arm. This is the so-called exploration process.

A variety of exploration strategies have been proposed for implementing MAB, including $\epsilon$-greedy~\cite{tokic2010adaptive}, Upper Confidence Bound (UCB)~\cite{li2010contextual}, Thompson Sampling (TS)~\cite{agrawal2013thompson}, EXP3~\cite{auer2002nonstochastic}, and so on. $\epsilon$-greedy or TS-based methods estimate the potential rewards through a random sampling from a posterior distribution, UCB-based methods assume that the potential payoffs should be the upper confidence bound of the reward distribution, while EXP3 algorithms compute the potential rewards with an exponential function.

In earlier MAB approaches, arms are commonly presumed to be independently from each other. Further studies built on top of the contextual MAB have strengthened the connections among different arms, such as the Linear UCB and Neural UCB based approaches. Linear UCB assumed a linear relationship between the feature of each arm and the corresponding reward~\cite{li2010contextual,bouneffouf2012contextual,li2017provably}, whereas Neural UCB further extended the linear feature mapping to non-linear mapping through neural networks~\cite{allesiardo2014neural,zhou2020neural}. In addition, previous studies often hypothesized a stochastic rewarding process using simple distributions such as Bernoulli distribution. Some recent studies have experimented an explicit modeling of the rewarding mechanism with more complex, assumption-free processes, such as Gaussian processes\cite{krause2011contextual} or variational inference\cite{blundell2015weight,graves2011practical,gal2016dropout,lakshminarayanan2017simple}.

\subsection{E\&E for Online Recommendation}
\citet{balabanovic1998exploring} formalized the E\&E trade-off problem in the context of personalized recommendation: whether to recommend an item with high uncertainty or to recommend the item known to match user interest that we have learnt so far. They showed that, despite with the expense of presenting users with sub-optimal recommendation results, the adoption of an exploration strategy can facilitate the convergence of model training. Another potential benefit is that such an strategy makes the recommender system easily adapt to the change user interest, which is relatively difficult for the exploitation-only based approaches.

Multi-armed bandit strategies such as $\epsilon$-Greedy and Upper Confidence Bound (UCB) have been adopted to understand the utility of exploration for recommender systems. \citet{shah2017practical} experimented with the $\epsilon$-Greedy strategy, in which we adopted the vanilla recommendation algorithm at the probability of 1-$\epsilon$, and explored randomly by choosing an arbitrary arm at the probability of $\epsilon$. ~\citet{nguyen2019recommendation} further tested the effectiveness of UCB-based exploration strategies for product recommendation, and the experimental results demonstrated its superior performance over other bandit strategies such as EXP3 and $\epsilon$-Greedy.

Despite the existing studies, the majority of E\&E algorithms for online recommendation systems generally follow the contextual multi-armed bandit modeling framework~\cite{li2010contextual}. \citet{li2010contextual} are the first to develop such an approach for personalized news recommendation in Yahoo! Homepage. To be specific, a contextual MAB algorithm selects an arm (i.e., by recommending an item to a user) at each step based on the policy that leverages the exploitation of items under the current knowledge, with the exploration of items with high uncertainty. Later studies have further extended this framework from various aspects and achieved significant improvements on their specific application contexts~\cite{li2011unbiased,vanchinathan2014explore,mcinerney2018explore,ban2021multi,zhang2020neural,song2021underestimation,guo2020deep,zheng2022implicit}.

In developing the E\&E algorithms, researchers commonly believe that high uncertainty is a good indicator of potential reward for exploration and a large body of work has been devoted to estimating uncertainty. For example, \citet{gal2016dropout} and \cite{blundell2015weight} attempted to approximate model uncertainty via Monte Carlo Dropout and Bayesian Neural Networks, respectively. \citet{song2021underestimation} adopted the gradient-based neural-UCB and neural Thompson Sampling for uncertainty estimation. \citet{du2021exploration} proposed a variational inference based approach called Deep Uncertainty-Aware Learning (DUAL), to estimate uncertainty with better accuracy. 

In summary, when developing a Multi-Armed Bandit (MAB) algorithm, the above studies mostly concentrated on estimating the potential reward for the selected arm (e.g., an item in a recommender system), whereas it does not take into account its subsequent effects on the recommendation service. We argue that such an effect can be non-trivial for an online recommendation system as the collected training data will be different after adopting a certain exploration strategy. To this end, we propose an Adversarial Gradient based Exploration approach to explicitly quantify such an effect, which will be described with more details in the below sections.

\section{Methodology} \label{sec:model}
In this section, we introduce our proposed \textbf{A}dversarial \textbf{G}radient based \textbf{E}xploration (AGE) approach for CTR prediction.

\subsection{Preliminary} \label{subsec:preliminary}
Before delving into model details, we first provide a formal description for the problem. Assuming that we have a data collection $\mathcal{D}$, which contains a set of data samples with input features $\mathcal{X}$, and the corresponding labels $y$, i.e. $\mathcal{D}=\{(\mathcal{X}_i, y_i)\}^N_{i=1}$. The goal of a CTR model, as shown in Equation \ref{eq:ctr}, is to learn a function $f(\cdot)$ that predicts the click label with high accuracy. In modern industrial systems, both of the dense features and sparse features are commonly encoded as feature embedding in the deep neural models ~\cite{zhou2018deep,zhou2019deep}. Accordingly, we separate our model parameters into two components -- the feature embedding $h(X)$ mapping from the data input $X$, and the model parameters for neurons $\theta$. Hereafter, we will use $h$ to denote the embedding parameters for simplicity.
\begin{equation}
\label{eq:ctr}
 \hat{y}_i=f(h(X_i), \theta)
\end{equation}

We further denote the exploitation-exploration process in a recommender system as follows. At each time step, an E{\&}E policy tries to recommend an item to a target user. The item is selected by considering the expected reward based on the current knowledge, and meanwhile allocating resources to explore items that the system has less knowledge of (i.e., items with large prediction uncertainty). In this way, the system might be better-off for cumulative rewards in the long run. In practice, the predicted CTR is often employed as the current expected reward, and the uncertainty is often obtained through Monte Carlo dropout~\cite{gal2016dropout}. Afterwards, exploration strategies such as UCB~\cite{auer2002using,li2010contextual} and Thompson Sampling (TS)~\cite{chapelle2011empirical} are utilized for the final ranking. Specifically, UCB adopts the upper confidence bound for exploration, whereas TS calculates the ranking score by sampling from the estimated distribution (with the predicted CTR as mean and prediction uncertainty as variance).

\subsection{Pseudo-Exploration for CTR Prediction}
\label{sec:model_preexp}
Conventional Exploration-Exploitation research mainly focuses on estimating prediction uncertainty, whereas the subsequent effect of exploration on model training is not properly considered. From the perspective of online learning~\cite{anderson2008theory}, the adoption of an exploration strategy also affects the collecting of training samples, which further influences model learning. Suppose that we have an item (along with the user) to be explored, and further assume that we will receive a user feedback $y^*$ if it is explored. With this new feedback, our model needs to minimize a new loss and updates model parameters. We define this process as one step of \textbf{pseudo-exploration}.

The primary goal of pseudo-exploration is to seek for the change of model parameters so that it can reflect model updating after an exploration. We believe this process mostly impacts the item (or user)-specific embedding, whereas only trivial adjustment is needed for non-embedding parameter $\theta$ as $\theta$ strives to accommodate for all of the data samples rather than a single item. Therefore, we keep $\theta$ intact, and focus on the updating of embedding $h$. To this end, we represent the above process using Equation \ref{pseudo-exploration}. Here, $\mathcal{L}(\cdot)$ denotes the loss, where the cross-entropy function is commonly adopted for CTR prediction. Moreover, we introduce the constraint $\|\Delta_h\|_2\leq \lambda$ to limit the maximum change of embedding.
\begin{align}
\label{pseudo-exploration}
    \Delta_h(\lambda, y^*) = \argmin_{\|\Delta h\|_2\leq \lambda} {\mathcal{L}}(f(h+\Delta_h|\theta), y^*)
\end{align}

With the Lagrange Mean Value Theorem, and upon the condition that the L2 norm of $\Delta_h$ approaches to zero, we can deduce the loss function (which is abbreviated to $\mathcal{L}(h+\Delta_h|\theta, y^*)$ for simplicity) to Equation \ref{eq:delta_y}. Placing it back to Equation \ref{pseudo-exploration}, the minimal value of the loss function is obviously on the situation where $\Delta_h$ has the opposite direction as $\nabla_h{\mathcal{L}(h|\theta,y^*)}$, and the scale equals to $\lambda$. This can be illustrated with the below Equation \ref{eq:gradient-solution1}.
\begin{equation}
\label{eq:delta_y}
\begin{split}
\mathcal{L}(h + \Delta_{h}|\theta, y^*)
    = \mathcal{L}(h|\theta, y^*) + \Delta_h \cdot \nabla_h{\mathcal{L}(h|\theta, y^*)}, \;\|\Delta_h\|_2\rightarrow{0}
\end{split}
\end{equation} 

\begin{equation}
\label{eq:gradient-solution1}
\Delta_h(\lambda, y^*)
      = -\lambda \cdot 
         \frac{\nabla_h{\mathcal{L}(h|\theta, y^*)}}
              {\|\nabla_h{\mathcal{L}(h|\theta, y^*)}\|_2}
\end{equation}

In practice, we often directly use the original gradient $\nabla_h{\mathcal{L}(h|\theta, y^*)}$ instead of the normalized gradient in Equation~\ref{eq:gradient-solution1}. By resolving the partial derivative with the chain rule, and further adopting the cross-entropy loss function, we are able to obtain the solution as shown in Equation \ref{eq:gradient-step1}. Here, we re-scale the hyper-parameter from $\lambda$ to $\lambda'$ to keep the equation stands. Even though they carry different meanings, we use them exchangeable hereafter because they are hand-tuned hyper-parameters.
\begin{equation} \label{eq:gradient-step1}
\begin{split}
\Delta_h(\lambda, y^*)
      &= -\lambda \cdot 
          \nabla_{f(h|\theta)}{\mathcal{L}(h|\theta, y^*)} \cdot 
          \nabla_h{f(h|\theta)} \\
      &= \lambda' \cdot
         (f(h|\theta)-y^*) \cdot
         \frac{\nabla_h{f(h|\theta)}}
              {\|\nabla_h{f(h|\theta)}\|_2} \\
\end{split}
\end{equation}

We further simplify the solution using Equation \ref{eq:gradient-solution}. Here, the normalized gradient $\vec{g}$ reflects the direction of the derivative of model output with respect to the input embedding. The difference between the predictive score and the true user feedback $f(h|\theta) - y^*$ is actually the difference between the prediction CTR and the real CTR in a probabilistic meaning, which will be represented by prediction uncertainty $\delta_y$ hereafter. Note that with the above transformation, the estimation of $\Delta_h$ no longer depends on the true user feedback $y^*$, which is unavailable beforehand. 
\begin{equation} \label{eq:gradient-solution}
\begin{split}
\Delta_h(\lambda, y^*)
       &= \lambda' \cdot \delta_y \cdot \vec{g}
\end{split}
\end{equation}

\begin{equation} \label{eq:adversarial-gradient}
\text{with}
\;\; \vec{g} = \frac{\nabla_h{f(h|\theta)}}
               {\|\nabla_h{f(h|\theta)}\|_2}
\;\;
\text{and}               
\;\; \delta_y = f(h|\theta)-y^*
\end{equation}

It is worth noting that the above Equation~\ref{eq:gradient-solution} is equivalent to finding $\Delta_h$, with the constraint of $\|\Delta_h\|_2\leq\lambda'\delta_y$, that maximizes the change of prediction output (Deduced the same as Equation \ref{pseudo-exploration} to Equation \ref{eq:gradient-solution1}.). This can be illustrated by Equation~\ref{eq:maximize-ftheta}, which shares the same form as adding an adversarial perturbation to the input~\cite{moosavi2017universal,goodfellow2014explaining,rozsa2016accuracy}. A detailed proof for their equivalence is provided in the Appendix material. For this reason, we treat $\vec{g}$ as the \textbf{Adversarial Gradient}, and name our approach as the Adversarial Gradient based Exploration. In addition to the uncertainty estimation, AGE moves one further step by redefining the utility of an exploration as its direct influence on model learning.

\begin{equation} \label{eq:maximize-ftheta}
\Delta_h(\lambda, y^*) = \argmax_{\|\Delta_h\|_2\leq \lambda'\delta_y} f(h+\Delta_h|\theta)
\end{equation}

Equation \ref{eq:maximize-ftheta} shows that an effective exploration in AGE should let the change of input embedding go towards the direction leading to the maximal change of prediction output (i.e. adversarial gradient $\vec{g}$), along with the strength of exploration measured by the prediction uncertainty (i.e. $\delta_y$). In this way, the exploration brings in a substantial adjustment to the prediction score. This also aligns with our expectation -- an exploration resulting little change is worthless because the model does not gain any new knowledge after the exploration.

After obtaining $\vec{g}$ and $\delta_y$, we compute the exploration-based model prediction $\hat{y}_e$ with Equation~\ref{eq:exp-predict}. This differs from the mainstream E\&E research, in which the final prediction is a direct summation of prediction score and uncertainty score. For example, in UCB-like approaches~\cite{auer2002using,li2010contextual}, the upper bound of prediction uncertainty is added to the prediction score for exploration. Our approach transforms the exploration problem into the change of input embedding, resulting in a more stable prediction distribution in practice. With the computed score of $\hat{y}_e$, our system will then rank items based on such a score. The exploitation-exploration trade-off is implicitly encoded as the amount of change for input embedding.

\begin{equation} \label{eq:exp-predict}
\hat{y}_e = f(h + \lambda'\cdot\vec{g}\cdot\delta_y | \theta)
\end{equation}

\begin{figure}[!t]
	\centering
	\includegraphics[width=1\linewidth]{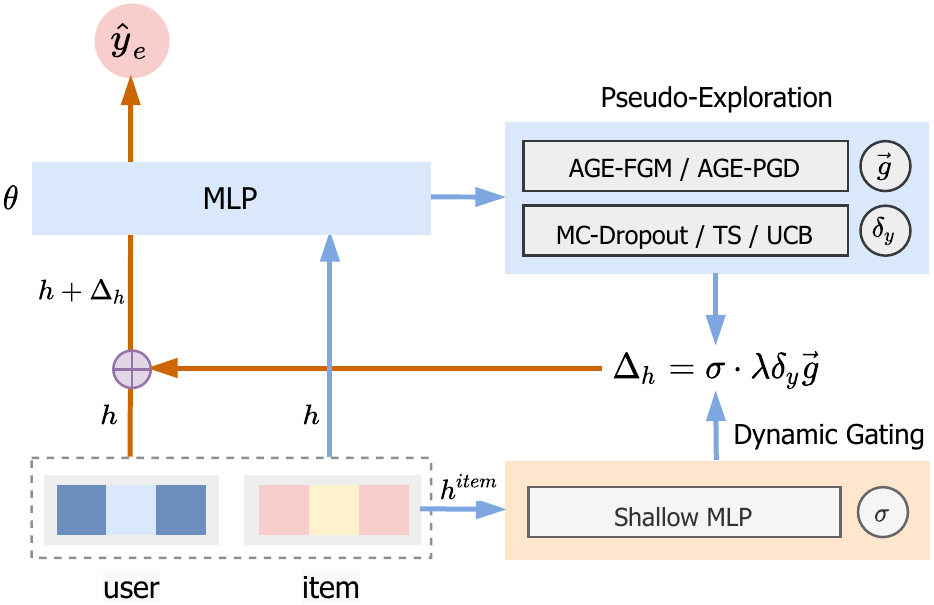}
	\caption{An illustration of the Adversarial Gradient-based Exploration (AGE) approach. It consists of three main components: a standard neural model for CTR prediction, a pseudo-exploration module, and a dynamic gating module.}
	\label{fig:age}
\end{figure}

\subsection{Parameter Computation}
In this section, we describe our approaches for computing the uncertainty $\delta_y$ and the adversarial gradient $\vec{g}$ in Equation~(\ref{eq:exp-predict}).

\subsubsection{Uncertainty.}
\label{sec:model_preexp_bl}
As shown in Equation~\ref{main_model_mc}, we adopt the commonly-used Monte Carlo Dropout (MC-Dropout) approach for uncertainty estimation ~\cite{gal2016dropout,blundell2015weight,song2021underestimation}. Here, $M$ stands for a mask matrix and its tensor shape aligns with $\theta$, $\odot$ represents the Hadamard product. Therefore, $M\odot\theta$ is equivalent to conduct a dropout on $\theta$. 
\begin{equation}
    \label{main_model_mc}
    \hat{y}=f(h | \theta \odot M)
\end{equation}

Here, we employ MC-Dropout for two reasons. First, it does not require training multiple models, which is particularly important for industrial systems since the production model training is highly resource-expensive. Second, MC-Dropout does not change model architecture, making it easy to be adapted in production systems.

By varying the mask matrix $M$, we are able to obtain different prediction scores. The model uncertainty $\delta_y$ will then be estimated from those predictions. For UCB based methods~\cite{auer2002using}, the variance of prediction scores is usually treated as the uncertainty (Equation ~\ref{eq_ucb}). Here, $M_i$ stands for one dropout setting, $f(h|\theta)$ denotes the predictions from the non-dropout model~\footnote{In practice, $f(h|\theta)$ can also be approximated by averaging all of the $N$ dropout models; however, we do not see much difference to a direct adoption of the non-dropout model for $f(h|\theta)$.} and we will conduct the dropout repeatedly for $N$ times.
\begin{equation}
\label{eq_ucb}
\delta_y = 
\sqrt{
    \sum_{i=1}^{N}
    \Big(
        \frac{(f(h|\theta \odot M_i)-f(h|\theta))^2}{N-1}
    \Big)
}
\end{equation}

With regards to the Thompson Sampling based approaches~\cite{chapelle2011empirical}, the uncertainty $\delta_y$ can be measured by the difference between a dropout model $f(h|\theta \odot M)$ and the non-dropout model $f(h|\theta)$, as shown in Equation~\ref{eq_ts}. According to Song et al. ~\cite{song2021underestimation}, this strategy can be thought as the sampling from a posterior distribution of CTR; therefore, Equation~\ref{eq_ts} is essentially the approximation of the Thompson Sampling approach in practice.
\begin{equation}
\label{eq_ts}
\delta_y = f(h|\theta \odot M) - f(h|\theta)
\end{equation}

\subsubsection{Adversarial Gradient.}
\label{sec:model_preexp_agm}
To obtain the normalized adversarial gradient $\vec{g}$, we examine two different approaches in this paper. At first, we adopt the Fast Gradient Method (FGM)~\cite{goodfellow2014explaining}, and approximate $\vec{g}$ through one-step update (Equation~\ref{eq:adversarial-gradient}). To improve the estimation performance, we further utilize the Project Gradient Descent (PGD)~\cite{2017Towards} approach, and update the gradient iteratively for $T$ steps. This can be illustrated by Equation~\ref{eq:PGD}.

\begin{equation}
\label{eq:PGD}
\begin{split}
    g^0 &= 0,\\
    g^{t} &= \vec{g}^{t-1} + \nabla_{h}f(h+g^{t-1}|\theta), \; g^{t} = \frac{g^{t}}{{\|g^{t}\|_2}},\\
    \vec{g} &= g^T.
\end{split}
\end{equation}

\subsection{Dynamic Gating Unit}
\label{sec:model_tu}
Under the conventional top-K recommendation paradigm, only a small number of highly effective items will be displayed to end users. With this background, we argue that exploring the items whose true click-through rates are low is ineffective. Through exploration, we may obtain more accurate predictions for those items; however, due to the noncompetitive prediction scores, they still cannot be displayed in the post-exploration stage. Accordingly, such an exploration should be avoided. This is particularly true for industrial systems, which cannot risk too much resource on exploration.

For this reason, we introduce a \textbf{Dynamic Gating Unit} (DGU), as illustrated by Equation~\ref{eq:score-gating-unit}, to control whether or not should we explore. Here, $\sigma$ denotes a zero-one gating function. In a highly-personalized system, click-through rate is determined not only by the item but also by the current user. Therefore, the dynamic gating unit should make the decision at the granularity of each user-item pair. In this paper, we adopt a simple heuristic for the gating unit -- if the prediction score of a user-item pair is larger than the item-level average of CTR, such an exploration should be encouraged; otherwise, it should be suppressed. We believe that this heuristic is only one type of design for the gating unit, and there are definitely many other alternatives. As for the gating function, one can also adopt other formats beyond the zero-one function. However, they are non-goals for this paper.
\begin{equation}
\label{eq:score-gating-unit}
\hat{y}_e =
    f(h + \sigma \cdot \lambda \cdot \delta_y \cdot \vec{g} | \theta) 
\end{equation}

The item-level CTR can be simply approximated through an average of historical click-through rates. For a better approximation performance, we go beyond this simple approach by developing a \textbf{shallow} DNN network that only utilizes item features (denoted by $h^{item}$). The Dynamic Gating Unit can then be represented by Equation~\ref{eq:gating-unit}. Here, $f(h|\theta)$ and $f_{dgu}(h^{item}|\theta_s)$ indicate the predicted CTRs from the main network and the shallow network, respectively. It is worth noting that the shallow network shares the same embedding parameters as the main prediction model; however, it does not participate the updating of embedding parameters during model training. This avoids the adverse effect on the main network from training the shallow network.
\begin{equation}
\label{eq:gating-unit}
\sigma = 
    \left\{
        \begin{matrix}
            1, &  f(h|\theta) \geq f_{dgu}(h^{item}|\theta_s)  \\
            0. &  f(h|\theta) < f_{dgu}(h^{item}|\theta_s)  \\
        \end{matrix}
\right. 
\end{equation}

\subsection{Overall Architecture}
Up to now, we have explained all of the components for the proposed Adversarial Gradient-based Exploration. To provide an overall picture, we illustrate its architecture in Figure~\ref{fig:age}. In addition to the standard neural CTR prediction model, AGE comprises two additional components: a Pseudo-Exploration module that simulates one-step of model training so that the exploration actions can facilitate future model learning, and a Dynamic Gating Unit (DGU) that helps prevent less effective explorations in practice.

By integrating the above modules, our final exploration-based click-through rate prediction model shall minimize the below loss function. Here, $\mathcal{L}$ denotes the standard cross entropy loss. The whole algorithm details are provided in Algorithm \ref{alg:1}.

\begin{equation} \label{loss} 
\mathbb{L} = \argmin_{\theta,\theta_s}\mathcal{L}(f(h|\theta\odot M),y^*) + \mathcal{L}(f_{s}(h^{item}|\theta_s),y^*) 
\end{equation} 

\begin{algorithm}[t] 
\normalsize 
\caption{Algorithm details for the proposed AGE approach.} 

\begin{flushleft} 
\hspace*{0.01in} 
{\bf Input:} Input feature $X$ and hyper-parameter $\lambda$.\\ 

\hspace*{0.01in} 
{\bf Output:} Exploration-based predictive CTR $\hat{y}_e$. 
\end{flushleft} 

\begin{algorithmic}[1] 
\State Compute the original predictive CTR $\hat{y}$ (Eq. \ref{eq:ctr}). 

\State Compute the uncertainty $\delta_y$ for UCB-like approaches (Eq. \ref{eq_ucb}) and TS-like approaches (Eq. \ref{eq_ts})

\State Compute the normalized adversarial gradient $\vec{g}$ (Eq. \ref{eq:adversarial-gradient} or Eq. \ref{eq:PGD} according to the adopted estimation approach)

\State Compute the dynamic gating unit (Eq. \ref{eq:gating-unit})

\State Compute the final score $\hat{y}_e$ (Eq. \ref{eq:score-gating-unit}).
\State \Return $\hat{y}_e$

\end{algorithmic}
\label{alg:1}
\end{algorithm}

\section{Data and Evaluation}
\label{exp}

\subsection{Datasets}
To understand the effectiveness of the proposed AGE approach, we firstly conduct a set of experiments with Yahoo! R6B dataset~\cite{li2011unbiased}. This dataset contains around 28 millions of user visits collected from the \textit{Today} Module of Yahoo! frontpage during a 15-day period of time in October 2011. Overall, Yahoo! R6B dataset contains 652 unique articles. For each visit, there are around 38 candidate articles (only partial of them were displayed to end users), along with the user click feedback information, are recorded. This enables us to evaluate an exploration strategy through replaying the recommendation process in the offline manner. Each data sample (i.e. each user visit) consists of the below information:

\begin{itemize}
\item A set of user features such as gender and age represented by 136-dimensional multi-hot vectors;
\item A set of candidate articles for recommendation. The identifiers for the displayed articles were recorded, and this article was chosen uniformly at random during online serving;
\item A 0/1 label indicating whether the displayed article was clicked by the user or not, i.e. the ground truth information.
\end{itemize}

In addition to the Yahoo! R6B dataset, we also evaluate our experiments with online A/B testing. Our production model is trained over billions of data samples in daily basis, and the model is deployed with the online learning paradigm. A detailed description regarding to the online dataset will be provided in Section \ref{sec:ind_exp}.

\subsection{Evaluation Metrics}
\label{subsection:evaluation-metric}
For the offline evaluation, we utilize the total number of user clicks as an approximation for the \textbf{cumulative rewards}. This aligns with most of the previous studies ~\cite{song2021underestimation,du2021exploration}, in which user click is often treated as exploration pay-off in personalized recommnder systems. A large number of user clicks usually indicate a better performance for an exploration strategy. 

With respect to the online A/B testing, we employ several standard metrics such as click-through rate (CTR) and prediction accuracy (e.g. PCOC) for evaluation. Here, PCOC (predicted CTR over the true CTR) examines whether the predictive score aligns with the actual click rate. For this metric, our goal is to obtain a value that is closer to 1. In the context of online advertising, we also evaluate the exploration strategy with a top-line business metric named AFR (Advertiser Follow-up Rate), which measures the willingness of an advertiser to renew its contract with our platform. An effective exploration can facilitate the growth of long tail advertisers, which can eventually improve such a metric. 

\subsection{Implementation Details}
We utilize the same neural architecture for the backbone CTR model across all exploration strategies, namely a three-layer MLP with 256, 64 and 2 nodes each. Particularly, in the AGE model, we adopt a 2-layer MLP for the Dynamic Gating Unit.

For model training, we employ the Adam optimizer~\cite{kingma2014adam} and the learning rate is set to 1e-5. With regards to AGE, we set the exploration step size $\lambda$ as 1e-3, and dropout rate as 0.01. If not mentioned explicitly, the PGD algorithm is applied for computing the adversarial gradient (Equation ~\ref{eq:PGD}). For UCB-based exploration strategies, the number of dropout times $N$ is set to 20 (Equation ~\ref{eq_ucb}). 

During offline experiments, we firstly train a standard CTR prediction model with 80,000 data samples (split by time). This model will then be used for warming up other predictive models so that the experiment models do not start from random predictions. All of our evaluations are conducted with the remaining samples. 
 
\subsection{Compared Methods}
\label{subsec:compared-method}
To understand the effectiveness of our proposed AGE approach, we include the below eleven baseline methods. They are selected either because of their state-of-the-art model performances, or because they are closely related to the idea of AGE.

\begin{itemize}
\item \textbf{DNN-vanilla strategy.} A pure click-through rate prediction model without any exploration strategy. This corresponds to the most common production practice. The deep CTR model is trained with the impression data and items are ranked according to their prediction scores. This will serve as the baseline for all of the other algorithms.

\item \textbf{Random strategy.} This strategy explores all of the items uniformly at random, and is served as a reference point.

\item \textbf{$\epsilon$-greedy strategy.} A simple multi-armed bandit exploration strategy which adopts DNN-vanilla method at the probability of 1-$\epsilon$, and explores randomly at the probability of $\epsilon$.

\item \textbf{Ensemble-TS and Ensemble-UCB.} These two methods train five deep CTR models using the same network structure. For Thompson Sampling, we randomly pick one model for serving. For UCB, we compute the variance of model predictions from the five models~\cite{murphy2012machine}.

\item \textbf{Gradient-TS and Gradient-UCB.} Same as Ensemble-TS and Ensemble-UCB, except that we replace the prediction variance with the L2 norm of the gradient~\cite{song2021underestimation,zhou2020neural}.

\item \textbf{GP-TS} and \textbf{GP-UCB.} These two methods utilize the Gaussian Process for estimating the prediction variance ~\cite{du2021exploration}.

\item \textbf{UR-gradient-TS} and \textbf{UR-gradient-UCB.} On top of Gradient-TS and Gradient-UCB, we further adopt the Underestimation Refinement methods for variance estimation~\cite{song2021underestimation}.
\end{itemize}

To understand the effectiveness of each component in AGE, we further consider the below setups for a number of ablation studies.

\begin{itemize}
\item \textbf{AGE-TS w/o $\delta_y$.} To examine the usefulness of $\delta_y$ in Equation ~\ref{eq:gradient-solution}, we simply replace it with a random value sampled from a Gaussian distribution.

\item \textbf{AGE-TS w/o $\vec{g}$.} We further experiment the removal of normalized gradient $\vec{g}$ in Equation ~\ref{eq:gradient-solution}.

\item \textbf{AGE-TS w/o DGU.} We also try to remove the Dynamic Gating Unit, and examine the utility of DGU.

\item \textbf{AGE-UCB w/o DGU.} Same as above, except the experiment is conducted on top of the AGE-UCB approach.
\end{itemize}

\section{Experimental Results}
\label{sec:aca_exp}
This section starts with evaluating the effectiveness of our proposed AGE approach. The experimental results are presented in Table~\ref{tab:main_exp}. Meanwhile, as shown in Table~\ref{tab:ablation_study} and Table~\ref{tab:ablation_study2}, we experiment with a number of ablation studies for better understanding the effectiveness of each component in AGE. All of the above experiments are conducted with the Yahoo! R6B dataset. Later on, we further deploy AGE in a large-scale online displaying advertisement system and report its performance in Section \ref{sec:ind_exp}.

\subsection{Overall Performance}
We first evaluate the performance of AGE and the baselines with the cumulative rewards (which is measured by the total number of clicks), and the results are provided in Table \ref{tab:main_exp}. Based on that, we have several important observations. 

\begin{table}[t]
\centering
	\caption{An evaluation of the cumulative rewards (mean ± std) for AGE and baselines. Here, the cumulative reward is measured by the total number of clicks, and 50\% indicates that each method only utilizes half of the training data.}
	\resizebox{\columnwidth}{!}{
    \begin{tabular}{ccccc}
        \toprule 
        Models      & \multicolumn{2}{c}{\bf 100\% Training Data} & \multicolumn{2}{c}{\bf 50\% Training Data} \\
        \cmidrule(lr){2-3} \cmidrule(lr){4-5}
        & \# of clicks & Imp.(\%) & \# of clicks & Imp.(\%)  \\
        \midrule 
        DNN-vanilla  & 39149.4$\pm$748.2  & - & 19345.6$\pm$432.3  & -   \\
        \midrule
        random              & 26212.8$\pm$129.2  & -33.04 $\downarrow$  & 13323.6$\pm$89.4  & -31.13 $\downarrow$     \\ 
        $\epsilon$-greedy   & 42540.2$\pm$1534.9 & 8.661 $\uparrow$   & 20384.6$\pm$583.9 & 5.371 $\uparrow$   \\
        Ensemble-TS         & 43429.0$\pm$1034.5 & 10.93 $\uparrow$   & 20372.2$\pm$476.3 & 5.307 $\uparrow$  \\ 
        Ensemble-UCB        & 31985.6$\pm$967.8  & -18.30 $\downarrow$  & 14872.0$\pm$418.7 & -23.13 $\downarrow$  \\ 
        GP-TS               & 44069.8$\pm$925.2  & 12.57 $\uparrow$   & 20376.8$\pm$504.3 & -5.330 $\downarrow$   \\ 
        GP-UCB              & 36191.8$\pm$964.2  & -7.555 $\downarrow$  & 18923.4$\pm$442.3 & -2.182 $\downarrow$   \\ 
        \midrule
        gradient-TS         & 46829.6$\pm$727.3  & 19.62 $\uparrow$   & 23227.2$\pm$727.3 & 20.07 $\uparrow$  \\ 
        gradient-UCB        & 37655.4$\pm$1265.1 & -3.816 $\downarrow$  & 17362.4$\pm$347.8 & -10.25 $\downarrow$  \\ 
        UR-gradient-TS      & 47539.6$\pm$808.8  & 21.43 $\uparrow$   & 22052.4$\pm$585.4 & 13.99 $\uparrow$  \\ 
        UR-gradient-UCB     & 41509.6$\pm$887.7  & 6.029 $\uparrow$   & 19476.8$\pm$492.4 & 0.678 $\uparrow$  \\ 
        \midrule
        AGE-TS& \textbf{50111.0$\pm$709.4} & \textbf{28.00} $\uparrow$    &  \textbf{24875.2$\pm$428.6}  & \textbf{28.58} $\uparrow$   \\ 
        AGE-UCB& \textbf{47873.6$\pm$1084.5} & \textbf{22.28} $\uparrow$  &  \textbf{23042.4$\pm$601.2}  & \textbf{19.11} $\uparrow$  \\
        \bottomrule 
\end{tabular}}
\label{tab:main_exp}
\end{table}

First of all, most of the exploration-based algorithms outperform the non-exploration DNN-vanilla method. This is consistent with previous studies~\cite{song2021underestimation,du2021exploration}, and indicates the necessity of developing an effective exploration strategy. In addition, baseline models built on top of the Thompson Sampling (TS) approach all outperform the UCB-based ones, proving that Thompson Sampling is a better strategy for incorporating model uncertainty~\cite{du2021exploration}. Among all of the baselines, UR-gradient-TS achieves the best performance among the TS-based models, and UR-gradient-UCB receives the best performance among the UCB-based models. Particularly, UR-gradient-TS outperforms the DNN-vanilla by 21.3\% on the cumulative payoff. 

More importantly, the AGE-based methods outperform all of the baselines, demonstrating the effectiveness of utilizing adversarial gradient for exploration. Specifically, AGE-TS and AGE-UCB outperform the strongest baselines, i.e., UR-gradient-TS and UR-gradient-UCB, by 5.41\% and 15.3\%, respectively. The best performed AGE-TS approach improves over the benchmark method by 28.0\%. It is worth noting that AGE-UCB exhibits a comparable performance to AGE-TS, whereas this is not the case for other approaches. For example, gradient-UCB significantly under-performs the gradient-TS. This again illustrates the robustness of our AGE model.

Finally, we provide a closer examination of the difference between AGE and the two strongest baselines. Gaussian Process based approaches, namely GP-TS and GP-UCB, mainly focus on an accurate estimation of prediction uncertainty. A superior performance of AGE over the GP-based methods demonstrates the utility of incorporating the adversarial gradient for exploration. Gradient-based methods, such as gradient-UCB, gradient-TS, UR-gradient-UCB and UR-gradient-TS, indeed employ the gradient information but only focus on its conversion to the uncertainty. Whereas we discover, in AGE, that a combination of gradient and uncertainty to simulate the future model training is a more effective approach for exploration.

\subsection{Ablation Study}
To achieve a better understanding of the proposed AGE approach, we conduct a number of ablation studies in this section.

\subsubsection{Effect of Training Data.}
We believe that a robust exploration strategy should be less sensitive to the amount of available training data. Therefore, we experiment to remove half of the training data, and see how would the model perform in this case. As shown in Table~\ref{tab:main_exp}, both AGE-TS and AGE-UCB exhibit a relatively stable performance after data reduction. Particularly, their improvements over DNN-vanilla remain at the same level. However, this leads to an obvious performance drop for most of the baseline models. For instance, UR-gradient-TS shows a +21\% increase of cumulative reward with the full data, whereas such an improvement decreases to +13\% while using half of the data. This clearly demonstrates the robustness of AGE-based approaches. For this reason, we believe that AGE could handle long-tailed items more effectively in practice.

\subsubsection{Effect of Each Module.}
Gradient $\vec{g}$, uncertainty $\delta_y$ and dynamic gating unit (DGU) are the three important components for AGE (see Equation \ref{eq:score-gating-unit}). To figure out the usefulness of different components, we conduct a set of ablation studies by discarding each of them from AGE. To be specific, we will experiment with the above four approaches mentioned in earlier Section~\ref{subsec:compared-method}. According to the results from Table~\ref{tab:ablation_study}, we find that all of the three modules have positive contributions to AGE. An elimination of either module would cause adverse effect on model performance.

\begin{table}[t]
\centering
\caption{An evaluation of the cumulative rewards (mean ± std) for AGE after the modification of each module. The improvement is computed over AGE-TS for TS-based methods, and over AGE-UCB for UCB-based methods.}
\begin{tabular}{lcc}
    \toprule
    Models                & \# of clicks        & Imp.(\%) \\
    \midrule
    AGE-TS                            & 50111$\pm$709.4   & - \\
    AGE-TS w/o uncertainty $\delta_y$ & 45677$\pm$1276.9  & -8.85\%$\downarrow$  \\ 
    AGE-TS w/o gradient $\vec{g}$     & 44638$\pm$1032.2  & -10.9\%$\downarrow$  \\
    AGE-TS w/o DGU    & 47387$\pm$828.6   & -5.44\%$\downarrow$  \\ 
    
    \midrule
    AGE-TS w/ threshold 0.02 for DGU  &   47328.2$\pm$925.1 & -5.55\%$\downarrow$ \\
    AGE-TS w/ threshold 0.01 for DGU  &   48263.8$\pm$969.8 & -3.69\%$\downarrow$ \\
    AGE-TS w/ threshold 0.005 for DGU &   47596.8$\pm$987.8 & -5.02\%$\downarrow$ \\
    
    \midrule
    AGE-UCB                           & 47874$\pm$1084.5  & - \\
    AGE-UCB w/o DGU   & 38327$\pm$1212.1  & -19.9\%$\downarrow$  \\ 
    \bottomrule 
\end{tabular}
\label{tab:ablation_study}
\end{table}

\begin{table}[t]
\centering
	\caption{An evaluation of the cumulative rewards (mean ± std) for different gradient computation algorithms.}
    \begin{tabular}{lcc}
        \toprule
        Gradient Computation Methods  & \# of clicks          & Imp.(\%) \\ 
        \midrule
        FGM (with AGE-TS)   & 48698.2 $\pm$ 1102.5  &  -\\ 
        PGD (with AGE-TS)   & 50111.0 $\pm$ 709.4   &  2.90\%$\uparrow$ \\ 
\bottomrule 
\end{tabular}
\label{tab:ablation_study2}
 \end{table}

It is worth mentioning that DGU plays a more important role in AGE-UCB than it does in AGE-TS. Without DGU, AGE-UCB exhibits a 19.9\% drop of performance, while AGE-TS only shows a 5.44\% decrease of performance. This is attributed to the characteristic of each algorithm. By adopting the upper bound of prediction variability, UCB is over-confident about the items with high uncertainty, whereas most of them may have relatively low click-through rates. For this reason, the dynamic gating unit can help pre-filter the low-quality items that are unnecessary to explore. 

As shown in Figure \ref{fig:age}, our DGU module develops a shallow neural model to determine the zero-one gating threshold (see Equation~\ref{eq:gating-unit}). In addition to the dynamic threshold, we can also utilize a fixed value. Here, we experiment with three fixed threshold values and report their performances in Table \ref{tab:ablation_study}. We can see that the best fixed-threshold approach, i.e., AGE-TS with DGU threshold set to 0.01, still underperforms AGE-TS by -3.69\% in terms of cumulative rewards. Moreover, while setting the threshold values to 0.02 or 0.005, we see a further drop of model performance. We also extensively hand-tune other threshold values, and do not find a better performance compared to 0.01. This again demonstrates the usefulness of DGU.

\subsubsection{Effect of Gradient Computation.}
In this section, we analyze the effectiveness of FGM and PGD -- the two gradient computation algorithms. As shown in Table~\ref{tab:ablation_study2}, PGD outperforms FGM by 2.9\% in terms of the cumulative rewards, indicating the necessity of utilizing a more accurate, multi-step gradient computation approach.

\subsection{Online A/B Testing}
\label{sec:ind_exp}
We also deployed AGE to one of leading e-commerce display advertising systems in the world. We conducted an online A/B testing for a period of time over one month in April 2021. This algorithm is now serving for the major production traffic in our system.

\subsubsection{Experiment Setup.} 
During online experimentation, we segment the traffic into buckets based on the unique user identifier. In this way, a user will be assigned exclusively to one bucket. Here, we do not directly compare the performance between an exploration bucket and a non-exploration bucket since any E\&E strategy will sacrifice short-term efficiency for long-term reward. However, obtaining the long-term effect is challenging in industrial systems. Instead, for a fair comparison, we construct a few \textbf{fair buckets}, and evaluate the performance over those buckets.

\begin{figure}[!t]
	\centering
	\includegraphics[width=0.85\linewidth]{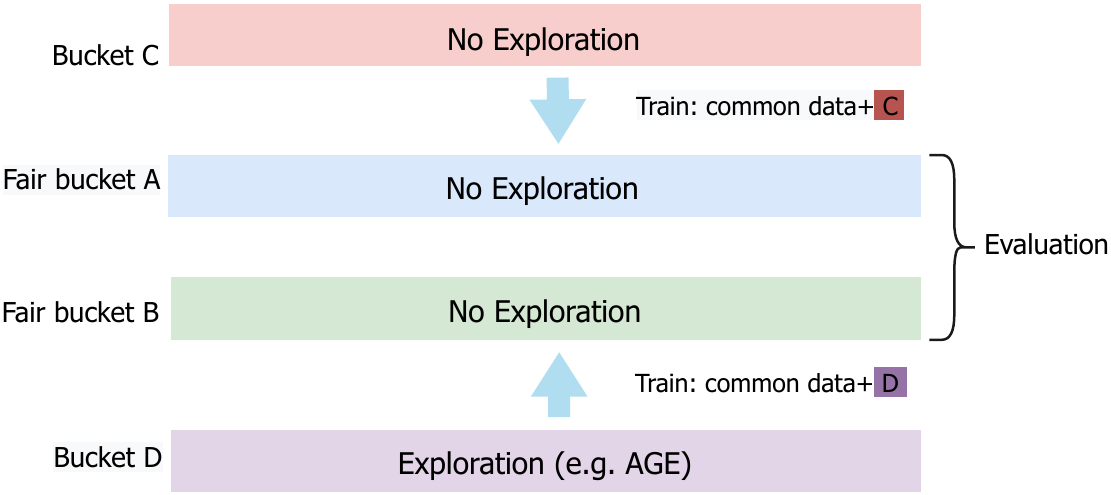}	\caption{An illustration of the design for fair buckets. Here, the user behavior data from Bucket C (D) will be used by Fair bucket A (B), respectively.}
	\label{fig:fair-bucket}
\end{figure}

The fair bucket is designed in the following way. As illustrated by Figure~\ref{fig:fair-bucket}, we first set up two buckets C and D with the same amount of traffic. Bucket D employs an exploration strategy such as AGE while bucket C uses the regular CTR model without any exploration strategy. Afterwards, we create two fair buckets A and B, both do not conduct any exploration. During model training, in addition to the common data (after the removal of data from all of four buckets), the model serving for bucket B utilizes the data from D and the model serving on bucket A receives data from C. Finally, we report the online performance for A and B.

\label{method}

In terms of model implementation, AGE inherits from our production model with a six-layer DIEN network~\cite{zhou2019deep}. The DGU module adopts a two-layer MLP structure. As for the exploration parameters, we set $N$ to 20 for UCB based strategies; and $\lambda$ to 0.002. In addition, the exploration is only conducted on items with the number of impressions fewer than 3,000. For online experimentation, we include Ensemble-TS and Ensemble-UCB as two baselines for simplicity. Here, the UR-gradient based approaches are not taken into account, which is due to the traffic limit, and we prefer to begin with the most basic exploration strategy for baseline.

\subsubsection{Evaluation Metrics.} 
As mentioned in Section \ref{subsection:evaluation-metric}, our online experiments are evaluated with several standard metrics, including the click-through rate (CTR), the total number of impressions for the exploring items (PV), and the PCOC (predicted CTR over the true CTR). We also include a top-line business metric named AFR to represent the satisfaction of advertisers.

\subsubsection{Experimental Results.}
Table \ref{table3} provides a comparison of model performance for the above-mentioned exploration strategies. AGE clearly outperforms all of the other methods -- it outperforms the production baseline by 6.4\% in CTR and 3.0\% in the number of impressions. Meanwhile, it also improves the prediction accuracy, i.e. the PCOC is much closer to 1. And more importantly, it improves the AFR metric by 5.5\%, indicating that our approach can even impact the experience of advertisers. Furthermore, as shown in Figure \ref{fig:ctr_cs}(b), we also discover that AGE can provide more accurate predictions even when the number of impressions are insufficient.

\begin{table}[t]
\centering
\caption{Online performance for different exploration strategies. Numbers indicate the improvement over the baseline.}
    \begin{tabular}{ccccccc}
        \toprule 
         Models      & CTR &PV & PCOC & AFR \\
        \midrule 
         Baseline & -&-&1.20&- \\
         Ensemble-UCB & -3.1\% & +0.2\% & 1.19&+0.7\%   \\
         Ensemble-TS  & +1.2\%  & +1.2\% &1.16&+2.3\%   \\
         AGE-TS  & \textbf{+6.4\%}  & \textbf{+3.0}\% &\textbf{1.10} &\textbf{+5.5\%}     \\
        \bottomrule 
\end{tabular}
\label{table3}
\end{table}

\section{Conclusion}
In this paper, we propose an Adversarial Gradient based Exploration (AGE for short) algorithm to deal with the Exploitation-Exploration problem for content recommendation. Different from most of the E\&E methods that concentrated on estimating the potential reward, our approach re-framed this problem in the data-driven context of online learning. More specifically, in addition to the prediction uncertainty of current model, AGE moves one further step by considering subsequent effect of exploration action on model training. This is achieved by deploying a pseudo-exploration module, in which we simulate the model updating process after an exploration action is conducted. Further analysis reveals that the prediction output of the updated model is equivalent to adding adversarial perturbations to input, which often improves the model robustness.

An E\&E strategy usually sacrifices short-term efficacy for long-term reward, making the industrial application a challenging topic. With regarding to the practical deployment issues, we propose a Dynamic Gating Unit to adaptively determine the value for item exploration. To understand the utility of our proposed AGE method, we conduct an extensive number of studies with both an academic dataset and an online A/B testing. Experimental results confirm the effectiveness of our proposed AGE-based exploration. 

Considering that an industrial recommender system often adopts a multi-stage cascading architecture, whereas we only apply AGE for the ranking stage in this paper. In the future, we shall extend this method to the other stages such as match or pre-rank stages.

\bibliographystyle{ACM-Reference-Format}
\bibliography{sample-sigconf}
\end{document}